# Negative and Positive Lateral Shift of a Light Beam Reflected from a Grounded Slab

Li-Gang Wang

*Institute of Optics, Department of Physics, Zhejiang University, Hangzhou, 310027, China*

Hong Chen

*Department of Physics, Tongji University, Shanghai 200092, China*

Nian-Hua Liu

*Department of Physics, Nanchang University, Nanchang 330047, China*

Shi-Yao Zhu

*Institute of Optics, Department of Physics, Zhejiang University, Hangzhou, 310027, China*

*and Faculty of Science and Technology, University of Macau, Macau*

We consider the lateral shift of a light beam reflecting from a dielectric slab backed by a metal. It is found that the lateral shift of the reflected beam can be negative while the intensity of reflected beam is almost equal to the incident one under a certain condition. The explanation for the negativity of the lateral shift is given in terms of the interference of the reflected waves from the two interfaces. It is also shown that the lateral shift can be enhanced or suppressed under some other conditions. The numerical calculation on the lateral shift for a realistic Gaussian-shaped beam confirms our theoretical prediction. ©2005 Optical Society of America.

It is well known that the Goos-Hänchen (GH) effect [1-2] refers to the lateral shift $\Delta$ of a totally reflected beam displaced from the path of geometrical optics. This phenomenon has been widely analyzed both theoretically [3-9] and experimentally [1, 10-14]. The investigation on the GH shift has been extended to other areas of physics, such as acoustics, quantum mechanics and plasma physics [15]. The GH shift is usually proportional to the penetration depth at the order of a wavelength. Many attempts have been made to achieve large lateral shifts (positive or negative) under different circumstances, such as absorbing media [4, 9], atomic resonant absorptive vapors [12, 16], resonant artificial structures [7, 17-18], negative-permittivity media [19-20], negatively refractive media [21-28], dielectric slabs [8, 29-30], multi-layered structures [6, 31-32]. Recently Li [8] found that the lateral shift of a light beam passing through a lossless transmitting slab could be negative due to the interference between the reflected waves from the slab's two interfaces. However, in Ref. 8, the amplitude of the reflected beam is strongly dependent on the parameters of the slab. In this paper, we consider a light beam reflecting from a dielectric slab backed by a metal, where the lateral shift of the reflected beam can also be negative with the advantage that the reflected beam intensity is almost equal to the incident one at any incident angles (this makes the comparison of the measured lateral shifts easy). When the thickness of the dielectric slab is suitable chosen, the negative lateral shift is dominated. The physics behind the negative lateral shift is the interference between the two reflected waves from two interfaces. At the same time, the lateral shift can also be positive, and it is shown that the lateral shift can be greatly enhanced or suppressed under certain conditions. Finally, we present a simulation for the lateral shift of a Gaussian-shaped beam.

Consider a TE-polarized light beam of angular frequency $\omega$ (corresponding wavelength $\lambda$) with angle $\theta$ incident from the vacuum ($\varepsilon_0 = 1$) upon a nonmagnetic dielectric slab backed by a metal ($\varepsilon_2$). The slab's thickness is $d$ and its dielectric constant is $\varepsilon_1$ ($\varepsilon_1$ is real number) as shown in Fig. 1. TM polarization can be discussed similarly. From Maxwell's equations and boundary conditions, the reflection coefficient could be expressed by [24-25]

$$r(\theta) = \frac{(k_{z0} - k_{z1})(k_{z1} + k_{z2}) + (k_{z0} + k_{z1})(k_{z1} - k_{z2})\exp(2ik_{z1}d)}{(k_{z0} + k_{z1})(k_{z1} + k_{z2}) + (k_{z0} - k_{z1})(k_{z1} - k_{z2})\exp(2ik_{z1}d)}, \quad (1)$$

where $k_{z0} = k_0 \cos\theta$, $k_0 = \omega/c = 2\pi/\lambda$, $k_{zi} = k_0(\varepsilon_i - \sin^2\theta)^{1/2}$ ($i = 1, 2$) when $\varepsilon_i > \sin^2\theta$, otherwise $k_{zi} = ik_0(\sin^2\theta - \varepsilon_i)^{1/2}$, and $c$ is the light speed in vacuum.

For the simplicity, we first consider the metal layer as a perfect conductor. In this case, Eq. (1) can be simplified into

$$r(\theta) = \frac{(k_{z0}^2 - k_{z1}^2) - (k_{z0}^2 + k_{z1}^2)\cos(2k_{z1}d) - i2k_{z0}k_{z1}\sin(2k_{z1}d)}{(k_{z0}^2 + k_{z1}^2) - (k_{z0}^2 - k_{z1}^2)\cos(2k_{z1}d)}, \quad (2)$$

and the phase of $r(\theta)$ can be analytically given by

$$\phi(\theta) = \tan^{-1}\left[\frac{-2k_{z0}k_{z1}\sin(2k_{z1}d)}{(k_{z0}^2 - k_{z1}^2) - (k_{z0}^2 + k_{z1}^2)\cos(2k_{z1}d)}\right]. \quad (3)$$

For the wide incident beam (i.e., with a very narrow angular spectrum, $\Delta k \ll k$), the lateral shift $\Delta$ of the reflected beam can be defined by: [8, 27] $\Delta = -\frac{\lambda}{2\pi\cos\theta}\frac{d\phi(\theta)}{d\theta}$. Then the analytical result of the lateral shift $\Delta$ is given by

$$\Delta = 2d\tan\theta \cdot \frac{\cos^2\theta + (\varepsilon_1 - 1)\frac{\sin[2k_{z1}d]}{2k_{z1}d}}{\cos^2\theta + (\varepsilon_1 - 1)\cos^2[k_{z1}d]}. \quad (4)$$

This is our key result of the present problem. For the optical denser medium ($\varepsilon_1 > 1$), the denominator in Eq. (4) is always positive. It is clear that when the inequality

$$\cos^2\theta + (\varepsilon_1 - 1)\frac{\sin[2k_{z1}d]}{2k_{z1}d} < 0 \quad (5)$$

holds, the lateral shift $\Delta$ becomes negative, i.e., the reflected beam will be appeared at the left side of the dotted line as shown in Fig. 1 (Ray 1). This result is counterintuitive in comparison with the prediction of geometric optics. In the frame of geometric optics, the lateral shift of the reflected wave from the first interface is zero and that of the reflected wave from the second interface is $2d\tan\theta'$ (where $\theta'$ is determined by Snell's law of refraction: $\sin\theta = \sqrt{\varepsilon_1}\sin\theta'$), see the dotted and dashed lines in Fig. 1. The inequality (5) indicates that the negative lateral shifts are much easily obtained at the large angles of incidence $\theta$ (or for the large value $\varepsilon_1$).

Figure 2 shows the dependence of $\Delta$ on the thickness $d$, where $\varepsilon_1 = 3.0$ (solid line) and $\varepsilon_1 = 9.0$ (dashed line) at wavelength $\lambda$, and the incident angle is $\theta = 80°$. It is shown that, the lateral shift can be negative, for example, it is equal to $-4.127\lambda$ at $k_{z1}d = 1.719$ for $\varepsilon_1 = 3.0$, and is even equal to $-4.681\lambda$ at $k_{z1}d = 1.638$ for $\varepsilon_1 = 9.0$.

In fact, we can rewrite Eq. (4) as

$$\Delta = \frac{2d\sin\theta\cos\theta}{\cos^2\theta + (\varepsilon_1 - 1)\cos^2[k_{z1}d]} + \frac{(\varepsilon_1 - 1)\tan\theta}{k_{z1}} \frac{\sin[2k_{z1}d]}{\cos^2\theta + (\varepsilon_1 - 1)\cos^2[k_{z1}d]}. \quad (6)$$

In this case, the first term is always positive and is proportional to the thickness $d$ by companying a periodical factor $\cos^2[k_{z1}d]$ with respect to $k_{z1}d$. It is the second term that leads the lateral shift to be negative due to the periodical function, $\sin[2k_{z1}d]$. The second term origins from the interference between two reflected waves from the two interfaces of the dielectric slab. The interference leads to the oscillation of the lateral shift [8]. However, in our case, the value of $|r(\theta)|$ is always equal to one, so that there is no angular shift even for the finite beam (which refers to the departure from the original propagation direction) that could not be avoided in Refs. [8, 29]. To average the first and second terms over $k_{z1}d$ in one period $[0, \pi]$ will erase the interference effect, and we can easily obtain the averaged lateral shift, $\bar{\Delta} = 2d \tan\theta'$, which is the same as the result of the reflected wave from the second interface, expected from the geometric optics.

Actually, under the condition of $k_{z1}d \ll \dfrac{\varepsilon_1 - 1}{2\cos^2\theta}$, the lateral shift is dominated by the second term in Eq. (6), and consequently the negative lateral shift can be observed much easier [see Fig. 3 (a)]. Conversely, when $k_{z1}d \gg \dfrac{\varepsilon_1 - 1}{2\cos^2\theta}$ is satisfied, the lateral shift will be dominated by the first term [see Fig. 3(b)], and consequently, $\Delta$ can be further approximated into $\Delta \approx \dfrac{d \sin 2\theta}{\cos^2\theta + (\varepsilon_1 - 1)\cos^2[k_{z1}d]}$. In this case, for the reflected beam, if $k_{z1}d$ satisfies the destructive resonant condition, $k_{z1}d = m\pi + \dfrac{\pi}{2}$ ($m$ is integer), $\Delta$ is enhanced and is approximately given by $\Delta_{\max} = 2d\tan\theta > \bar{\Delta}$ (because of $\theta > \theta'$). When $k_{z1}d$ satisfies the constructive resonant condition, $k_{z1}d = m\pi$, the lateral shift is greatly suppressed and it can be approximately expressed by $\Delta_{\min} = \dfrac{2d\sin\theta\cos\theta}{\varepsilon_1 - \sin^2\theta} = 2d\tan\theta' \dfrac{\tan\theta'}{\tan\theta} < \bar{\Delta}$ [see Fig. 3(b)]. The physical origin of the enhancement (or suppression) of the lateral shift comes from the destructive interference (or the constructive interference) of the reflected wave at the first interface with the incident wave.

In practice, the plane wave assumed above is an approximation. Normally the front of an incident wave is finite, that is to say, it has a profile with a certain width, typically a Gaussian profile. Therefore, we consider the lateral shift of the finite beam, such as a Gaussian-shaped beam, incident upon the system, for example, a glass layer backed by silver. The electric field of the incident beam at the plane of $z = 0$ is given by the following form:

$$E_y(x, z=0) = \frac{1}{\sqrt{2\pi}} \int A(k_x) \exp(ik_x x) dk_x, \tag{7}$$

where $A(k_x) = \frac{W_x}{\sqrt{2}} \exp[-\frac{W_x^2(k_x - k_{x0})^2}{4}]$ is the angular spectrum of the Gaussian-shaped beam with the angle $\theta$, and $k_{x0} = k_0 \sin\theta$, $W_x = W \sec\theta$, $W$ is the half-width of the beam at waist. The reflection coefficient $r(k_x)$ can be obtained from Eq. (1). Then the electric field of the reflected beam can be written by

$$E_y(x, z = 0) = \frac{1}{\sqrt{2\pi}} \int A(k_x) r(k_x) \exp(ik_x x) dk_x. \tag{8}$$

In the numerical simulations using Eqs. (7,8), we take $W = 13\lambda > \lambda$ (the incident beam is well collimated and $A(k_x)$ is sharply distributed around $k_{x0}$), and consequently, it is expected that the shape of the reflected beam is the same as that of the incident beam without obvious distortion. We use the peak position difference between the incident and reflected beams to denote the lateral shift $\Delta$, see the black dots in Fig. 4. In Fig. 4, we also plot the result from Eq. (4) by the thin lines. It is clear that our simulation result is in good agreement with our analytical result. From Fig. 4(a), with slab thickness $d = \lambda$, the lateral shift can be negative at large angle [also see the inset of Fig. 4(b)]. In addition, for slab thickness, $d = 6\lambda$, $\Delta$ can enhanced or suppressed at certain angles, which can be see in Fig. 4(b) by comparing with $\overline{\Delta}$ (the thick line).

In conclusion, the GH shift of a light beam, reflecting from a dielectric slab backed by a metal, is theoretically investigated. The analytic expression for the lateral shift by the stationary-phase method is obtained, and the numerical simulations confirm the analytical predictions. It is found that the lateral shift of the reflected beam can be negative as well as positive. The negative lateral shift is due to the interference of the reflected waves from the dielectric-air and dielectric-metal interfaces. This is a counterintuitive phenomenon in comparison with the prediction of geometrical optics. Meanwhile, the positive lateral shift can be enhanced or suppressed under certain conditions. It should be pointed out that, because the modulus of the reflection coefficient is equal to one, the angular shift is never appeared in our case and the shape of the reflected beam is nearly without distortion in our simulation.

This work was supported by National Natural Science Foundation of China (under contract No. 10547138) and RGC (HKBU2027/04P). Li-Gang Wang's e-mail address is sxwlg@yahoo.com.cn.

# Figure Captions

FIG. 1. Sketch of a light beam incident on a dielectric slab backed by a metal. The dotted and dashed lines are, respectively, the paths of reflection from the first and second interfaces, expected from geometric optics. Ray 1, the negative lateral shift, and Ray 2, the positive lateral shift of the reflected beam.

FIG. 2. Dependence of the lateral shift $\Delta$ on the thickness $d$ (rescaled by $k_{z1}d$) for the slab of $\varepsilon_1 = 3.0$ (solid line) and that of $\varepsilon_1 = 9.0$ (dashed line) at wavelength $\lambda$ with the angle $\theta = 80°$.

FIG. 3. Dependence of the lateral shift $\Delta$ on the thickness $d$ (rescaled by $k_{z1}d/\pi$) of the slab with $\varepsilon_1 = 3.0$ and incident angle $\theta = 85°$, under the conditions: (a) $k_{z1}d \ll \frac{\varepsilon_1 - 1}{2\cos^2\theta}$ and (b) $k_{z1}d \gg \frac{\varepsilon_1 - 1}{2\cos^2\theta}$. The first, second terms, and exact result of Eq. (6) are denoted by dashed, short dashed and solid lines, respectively.

Fig. 4. Simulation result (black dot) and analytical result (thin line) of the lateral shift for the slab thickness: (a) $d = \lambda$ and (b) $d = 6\lambda$. Inset of (b) shows the details near large angles. Thick line denotes the shift $\overline{\Delta}$ from the geometric optics. The parameters of the system are $\varepsilon_1 = 4.0$ and $\varepsilon_2 = -50 + 0.5i$.

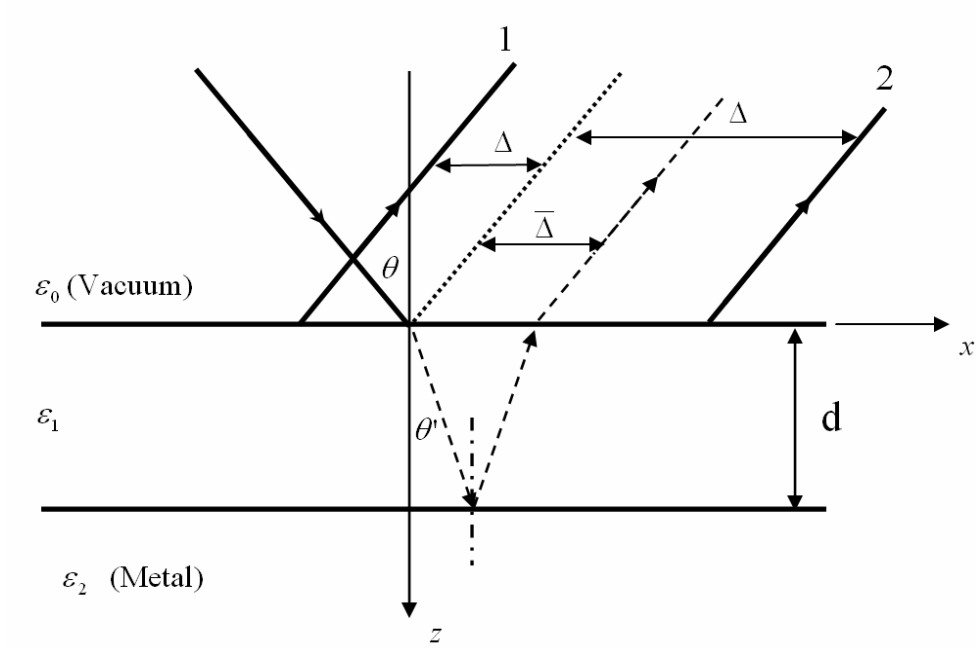

**FIG. 1.**

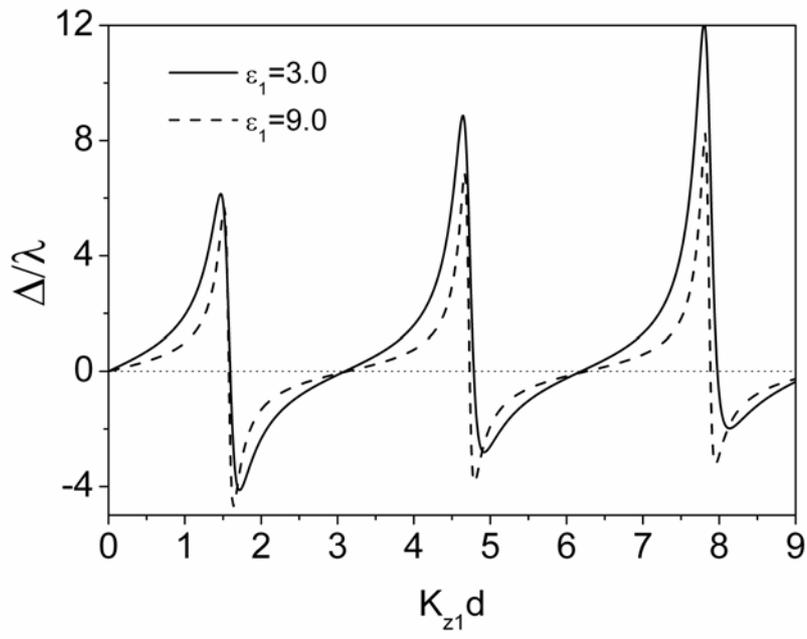

FIG. 2.

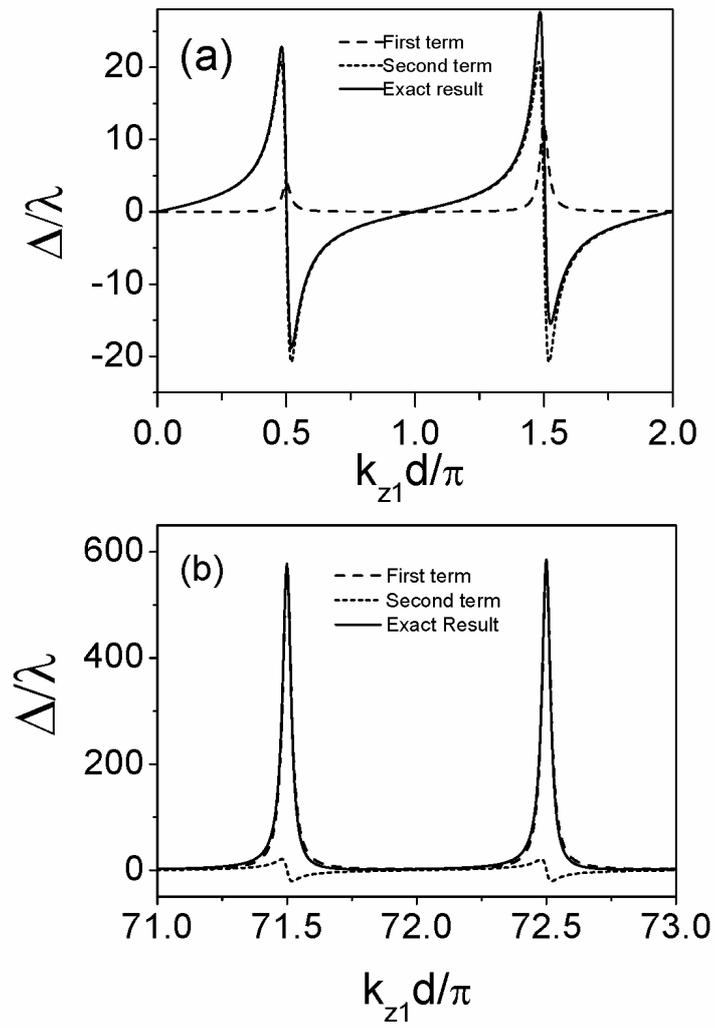

**FIG. 3**

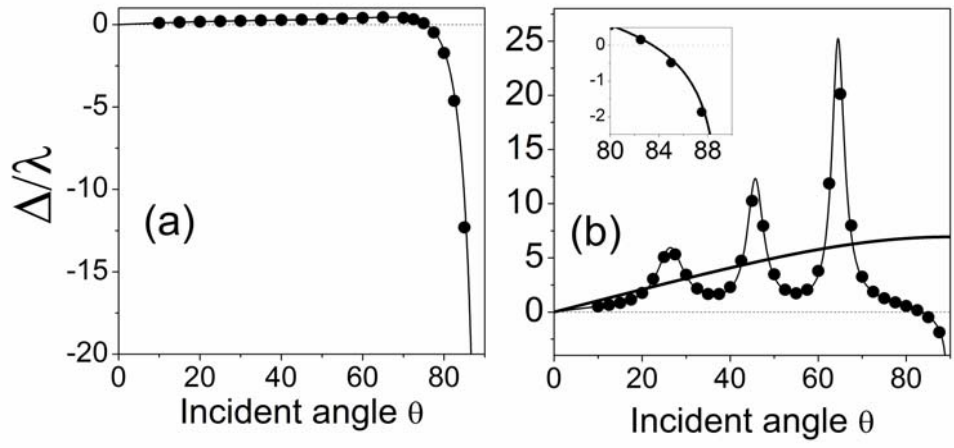

**Fig. 4.**